\documentclass{ifacconf}

\usepackage{graphicx, subfigure}      % include this line if your document contains figures
\usepackage{natbib}        % required for bibliography
\usepackage{subfigure}
\usepackage{amsmath}
\usepackage{amssymb}
\usepackage{comment}
\usepackage{xcolor}
\begin{document}
\begin{frontmatter}

\title{%Topography-induced changes in wildfire propagation in a physics-based advection-diffusion-reaction model./Impact of a topography extension on the dynamics of an advection-diffusion-reaction model for wildfire propagation. ??
Impact of topography and combustion functions on fire front propagation in an advection-diffusion-reaction model for wildfires
}
%Style for Full Contributions: Use Title Case for
  %Paper Title\thanksref{footnoteinfo}} 

%\thanks[footnoteinfo]{Sponsor and financial support acknowledgment
%goes here. Paper titles should be written in uppercase and lowercase letters, not all uppercase.}

\author[BS]{Luca Nieding} 
\author[BS]{Cordula Reisch} 
\author[BS]{Dirk Langemann}
\author[Four]{Adrián Navas-Montilla}

\address[BS]{Institute for Partial Differential Equations, TU Braunschweig, Germany (e-mail: l.nieding@tu-bs.de, c.reisch@tu-bs.de, d.langemann@tu-bs.de)}
\address[Four]{Fluid Dynamics Technologies Group, Aragon Institute of Engineering Research (I3A), University of Zaragoza, Spain (e-mail: anavas@unizar.es)}

\begin{abstract}                % Abstract of not more than 300 words. 

Given the recent increase in wildfires, developing a better understanding of their dynamics is crucial. For this purpose, the advection-diffusion-reaction model has been widely used to study wildfire dynamics. In this study, we introduce the previously unconsidered influence of topography through an additional advective term. Furthermore, we propose a linear term for the combustion function, comparing it with the commonly used Arrhenius law to offer a simpler model for further analysis. Our findings on the model's dynamics are supported by numerical simulations showing the differences of model extensions and approximations. 

\end{abstract}


\end{frontmatter}
%===============================================================================

\section{Introduction}

During the last twenty years, the frequency and intensity of extreme wildfires has increased, \cite{cunningham_increasing_2024}. This leads to a higher air pollution, and more housing areas are endangered by fires. 
Predicting the dynamical behavior of fires is therefore a crucial task to plan possible evacuation strategies. 
Mathematical models provide a basis for reliable tools. 
One class of mathematical models for wildfire spread are advection-diffusion-reaction (ADR) models describing the time evolution of the temperature and the quantity of combustible biomass.
Such a model was studied in \cite{asensio_wildland_2002}, followed by either numerical investigations e.g. in \cite{burger_exploring_2020} or mathematical analysis of travelling wave solutions in \cite{babak_effect_2009} and \cite{mitra_studying_2024}. 
Certain submodels and their interplay were studied both, analytically and numerically, in \cite{reisch_analytical_2024}. 
Based on these investigations, we study here the influence of the topography on the wildfire spread, which was introduced as well in \cite{burger_exploring_2020}.
The topography of the terrain influences the travelling wave speed and the direction of spread in an advection-like term. Due to approximations in the physical derivation of the mathematical model, the topographical influence may lead to non-physical behavior, which we overcome here by an additional correction term. 

The paper is structured as follows: We first derive the mathematical model based on physical mechanisms. Based on the model, we discuss the combustion function and the modeling of the topography influence on the advection. Afterward, we show numerical simulations that highlight the modeling choices and some undesired non-physical behavior due to topography. Finally, a correction term for the topography advection is introduced and validated. 

\section{Advection-diffusion-reaction model}
We start with the derivation of the model and then, we highlight certain mechanisms in more detail. 

\subsection{Mathematical Model}

We consider a two-dimensional spatial domain $\Omega \subset \mathbb{R}^2$ that models a region on a map where we are interested in the spread of the fire. 
For deriving the governing physical equations of the partial differential equation model, we consider a control volume $V = S \times [0, \ell]$ with a spatial height $\ell$ and the two-dimensional area $ S \subset \Omega $, which is a subset of the region of interest $\Omega$. 
The spatial vector $\mathbf{x} = (x_1,x_2)^\mathrm{T} \in \Omega$ describes a point on the map. 
The height of the control volume is introduced for the physical derivation, but it will be reduced in the later model. 

We assume a conservation of energy that is given by
\begin{equation}
    \frac{\mathrm{d}}{\mathrm{d}t} U  + \dot{Q}_\mathrm{adv} = \dot{Q}_\mathrm{cond} - \dot{Q}_\mathrm{conv} + \dot{Q}_\mathrm{rad} + \dot{Q}_\mathrm{reac},
    \label{conservationSingleTerms}
\end{equation}
where $U$ denotes the internal energy, and $\dot{Q}$ the heat fluxes by advection $\dot{Q}_\mathrm{adv}$, conduction $\dot{Q}_\mathrm{cond}$, convection $\dot{Q}_\mathrm{conv}$, radiation $\dot{Q}_\mathrm{rad}$ and reaction $\dot{Q}_\mathrm{reac}$. 

Integration of the combustible biomass over the height $z \in (0, \ell)$ gives a density of the combustible biomass $\rho(\mathbf{x},t)$ for each surface point $\mathbf{x} \in S $ and every time $t>0$, that is related as well to the mass of fuel in the control volume
\begin{equation}
    \rho(\mathbf{x},t) = \frac{m_\mathrm{f}(\mathbf{x},t)}{V}, \quad \rho_0(\mathbf{x}, t = 0) = \frac{m_{\mathrm{f}0}(\mathbf{x})}{V},
\end{equation}
where $m_\mathrm{f}(\mathbf{x},t)$ is the fuel mass at any time $t$ and $m_{\mathrm{f}0}(\mathbf{x})=m_\mathrm{f}(\mathbf{x},0)$ is the initial fuel mass.

The parameter $c \in \mathbb{R}_+$ gives the specific heat of the biomass and is assumed to be constant in time and space. 
The value $T(\mathbf{x},t)$ gives the temperature at a spatial point $\mathbf{x} \in \Omega \subset \mathbb{R}^2$ and a time $t\geq 0$. 
The temperature $T$ is one of the two variables in the upcoming model for spreading wildfires. 
The second variable is defined via the fraction of the fuel mass for any time and at the initial time, so 
\begin{equation}
    Y(\mathbf{x},t) = \frac{ m_\mathrm{f}(\mathbf{x},t)}{m_{\mathrm{f}0}(\mathbf{x})}.
\end{equation}

The change of the internal energy is then given by 
\begin{equation}
    \frac{\mathrm{d}}{\mathrm{d}t}U(\mathbf{x},t) = \frac{\mathrm{d}}{\mathrm{d}t}\int_V c \rho(\mathbf{x},t) T(\mathbf{x},t) \mathrm{d}V.
    \label{eq:innerE}
\end{equation}

We introduce the different heat fluxes that form together the physical model. 
The advection heat flux describes the amount of energy that leaves the control volume $V$ over the boundaries, e.g. due to wind, giving a directed heat transport. 
In formula, this is
\begin{equation}
    \dot{Q}_\mathrm{adv} =\int_{\partial V} c \rho(\mathbf{x},t)  T(\mathbf{x},t) \mathbf{v}(\mathbf{x},t) \cdot \mathbf{n} \, \mathrm{d} \Gamma, 
    \label{eq:AdvQ}
\end{equation}
with an advection velocity field $\mathbf{v}(\mathbf{x},t)$ and an outer normal vector $\mathbf{n}$ on the surface $\partial V$ with the infinitesimal area $\mathrm{d} \Gamma$. 
The conduction heat flux is given by diffusive heat transport with the thermal conductivity parameter $k$,
\begin{equation}
    \dot{Q}_\mathrm{cond} = \int_{\partial V} k \nabla T(\mathbf{x},t) \cdot \mathbf{n}\, \mathrm{d}\Gamma.
\label{eq:LeitQ}
\end{equation}
The convective heat flux is given by Newton's cooling law, 
\begin{equation}
    \dot{Q}_\mathrm{conv} = \int_S \alpha_0 (T(\mathbf{x},t)-T_\infty)\, \mathrm{d}\Gamma,
\label{eq:KonvQ}
\end{equation}
where $\alpha_0$ is the coefficient of the heat transfer and $T_\infty$ the ambient temperature of the air, soil or non-inflammable material inside the volume $V$. 

The radiation is given by the Stefan-Boltzmann law
\begin{equation}
    \dot{Q}_\mathrm{rad}  = \int_{\partial V}  \epsilon \sigma( T(\mathbf{x}+\delta,t)^4  -T(\mathbf{x},t)^4 ) \, \mathrm{d} \Gamma,
    \label{eq:radfirst}
\end{equation}
with the optical path length $\delta$, an emissivity factor $\epsilon$ and the Boltzmann constant $\sigma$. Using the Rosseland approximation gives the simplified expression
\begin{equation}
    \dot{Q}_\mathrm{rad} \approx \int_{\partial V} 4 \epsilon \sigma \delta \, T^3 \nabla T \cdot \mathbf{n} \, \mathrm{d}\Gamma,
    \label{eq:StrQ}
\end{equation}
which can be interpreted as a non-linear diffusion term.

The reaction heat flux models the main combustion process of the fire. 
The biomass is consumed during the combustion process, leading to a temporal change of the bio mass fraction
\begin{equation}
    \frac{\partial Y(\mathbf{x},t)}{\partial t} = - \Psi(T) Y(\mathbf{x},t),
    \label{eq:YDGL}
\end{equation}
with some combustion function $\Psi(T)$ that will be discussed later. 

The reaction heat flux then reads 
\begin{equation}
    \dot{Q}_\mathrm{reac} = \int_V - \mathcal{H} \rho_0 \frac{\partial Y(\mathbf{x},t)}{\partial t} \, \mathrm{d}V = \int_V \mathcal{H} \rho_0 \Psi(T) Y(\mathbf{x},t) \, \mathrm{d}V,
    \label{eq:ReakQ}
\end{equation}
where $\mathcal{H}$ is the combustion heat per unit mass. 

Combining all terms, we can formulate the physical full model in the reference volume,
\begin{equation*}
\begin{split}
   & \frac{\mathrm{d}}{\mathrm{d}t}\int_V c \rho(\mathbf{x},t) T(\mathbf{x},t) \mathrm{d}V + \int_{\partial V} c \rho(\mathbf{x},t)  T(\mathbf{x},t) \mathbf{v}(\mathbf{x},t) \cdot \mathbf{n} \, \mathrm{d} \Gamma  \\ & = \int_{\partial V} k \nabla T(\mathbf{x},t) \cdot \mathbf{n}\, \mathrm{d}\Gamma - \int_S \alpha_0 (T(\mathbf{x},t)-T_\infty)\, \mathrm{d}\Gamma  \\& +\int_{\partial V} 4 \sigma \epsilon \delta \, T^3 \nabla T \cdot \mathbf{n} \, \mathrm{d}\Gamma + \int_V \mathcal{H} \rho_0 \Psi(T) Y(\mathbf{x},t) \, \mathrm{d}V.   
\end{split}
\end{equation*}
Assuming a constant fuel mass density $\rho (t, \mathbf{x}) = \rho_0$, applying Gauss's divergence theorem, and reordering integration and differentiation leads to
\begin{equation*}
\begin{split}
    \rho_0 c &\int_V \left( \frac{\partial T(\mathbf{x},t)}{\partial t} + \nabla \cdot \left ( T(\mathbf{x},t) \mathbf{v}(\mathbf{x},t) \right ) \right) \, \mathrm{d}V \\ = &\int_V \nabla \cdot (4 \sigma \epsilon \delta T(\mathbf{x},t)^3 + k) \nabla T(\mathbf{x},t) - \alpha (T(\mathbf{x},t)-T_\infty)\\
    &+ \rho_0 \mathcal{H} \Psi(T) Y(\mathbf{x},t) \, \mathrm{d}V,
\end{split}
\end{equation*}
where $\alpha = \alpha_0 \ell$ is the coefficient of heat transfer.
As the control volume $V$ was chosen arbitrarily, we can follow that the partial differential equation system 
\begin{align} 
\begin{aligned}  \label{eq:finalModel}
    \rho_0 c\left(\frac{\partial T}{\partial t} + \nabla \cdot ( T \mathbf{v} ) \right) 
    &= && \nabla \cdot (k_t(T) \nabla T) - \alpha (T-T_\infty) \\
   & &&+ \rho_0 \mathcal{H} \Psi(T) Y   \\
         \frac{\partial Y}{\partial t} &= &&- \Psi(T) Y, 
    \end{aligned}
\end{align}
yields pointwise. 
Here, the diffusive mechanisms are combined to a function 
\begin{equation}\label{eq:nonlinear_diffusion}
    k_t(T)= 4 \sigma \epsilon \delta T^3 + k,
\end{equation}
including both, the linear diffusion by conduction and the non-linear diffusion by radiation. 

The resulting system is an advection-diffusion-reaction equation including multiple physics-based mechanisms for advection and diffusion. The reaction function for combustion couples the two equations. 
The influence of the basic mechanisms on the total system behavior was studied in \cite{reisch_analytical_2024} numerically and partly analytically. 

To summarize, the reaction, diffusion and advection mechanisms interfere with each other giving travelling wave solutions in meaningful parameter regimes. 

Here, we study different reaction functions $\Psi(T)$, and extend the model by a term for the influence of the topography of $S$ on the advection term.
This additional advection becomes relevant because the derivation of the model is based on a three-dimensional volume, but the processes are modeled only on a two-dimensional surface. 
Therefore, energy exchange in the third dimension needs to be modeled separately. 

Our numerical studies are based on the numerical schemes presented in \cite{reisch_analytical_2024}, where high order finite volume schemes with some operator splitting were used. 

\begin{table}[]
\caption{Model parameters used for numerical simulations, based on \cite{reisch_analytical_2024}.}
\begin{tabular}{llll}
Parameter                           & Symbol          & Value & Unit               \\ \hline
Fuel mass density                   & $\rho_0$        & 4           & kg m$^{-3}$           \\
Specific heat                   & $c$             & 1          & kJ kg$^{-1}$ K$^{-1}$ \\
Pre-exponential factor            & $A$             & 0.05                    & s$^{-1}$              \\
Combustion heat per unit mass & $\mathcal{H}$             & 5000        & kJ kg$^{-1}$          \\
Coefficient of heat transfer           & $\alpha$        & {[}0.1,10{]}           & kW m$^{-3}$ K$^{-1}$  \\
Ambient air temperature                      & $T_\infty$      & 300                    & K                     \\
Ignition temperature               & $T_\mathrm{pc}$ & 400                    & K                     \\
Activation temperature               & $T_\mathrm{ac}$ & 400                    & K                     \\
Thermal conductivity                  & $k$             & {[}0,4{]}              & kW m$^{-1}$ K$^{-1}$  \\
Stefan-Boltzmann-constant         & $\sigma$        & 0.057                     & $\mu$W m$^{-2}$ K$^{-4}$ \\
\end{tabular}
\\
\label{tab:modelParams}
\end{table}

\subsection{Model parameters and boundary conditions}

The partial differential equation system \eqref{eq:finalModel} is completed with initial conditions and boundary conditions. 
For simulation purposes and under the assumption that the fire does not reach the boundaries of the domain, we assume periodic boundary conditions. 
Let the two-dimensional computational domain be $\Omega = [a_1, b_1] \times [a_2, b_2]$. 
Then, the periodic boundary conditions read
\begin{equation}
    \left\{\begin{array}{ll} 
        &T((x_1,a_2),t) = T((x_1,b_2)),t), \hfill x_1 \in [a_1,b_1] \\ 
      & T((a_1,x_2),t) = T((b_1,x_2)),t), \quad x_2 \in [a_2,b_2]\\
        &   Y((x_1,a_2),t) = Y((x_1,b_2)),t), \quad t>0\\
        & Y((a_1,x_2),t) = Y((b_1,x_2)),t).  \\\end{array}\right.
        \label{eq:boundaryfinalmodel}
\end{equation}

For the initial conditions, we fix the initial biomass fraction to be one, due to the definition of $Y$.
The initial temperature should be at some $\mathbf{x} \in \Omega$ above the ignition temperature for starting a fire. 
More precise, we use the initial conditions 
\begin{equation}
   \begin{array}{ll} & T(\mathbf{x},0) = T_0(\mathbf{x}), \quad Y(\mathbf{x},0) = 1, \, \mathbf{x} \in \Omega 
    \\\end{array}
        \label{eq:initialfinalmodel}
\end{equation}
with $T_0(\mathbf{x})$ being $T_\infty$ for large parts of the domain and above $T_\mathrm{pc}$ for some regions. 

The model parameters are fixed for our simulations and given in Table~\ref{tab:modelParams}. 
We aim to focus on the coupling of the reaction term and the advection term. Therefore, we simplify the non-linear diffusion in Eq.~\eqref{eq:nonlinear_diffusion} to a linear diffusion by not considering the radiation effects. 

\subsection{Dynamics of the ADR-model in one spatial dimension}

 For gaining an understanding of the solution dynamics, we investigate first the ADR-model in one spatial dimension. 
 Therefore, we fix the computational domain $\Omega= [0,500]$, and therein initial conditions
\begin{equation}
    T_0(x) =  \left\{\begin{array}{ll}  400\, \mathrm{K}, & x \in [250-r,250+r] \\ 
        300 \, \mathrm{K}, & \mathrm{else}, \end{array}\right.
    \label{eq:initialCondition}
\end{equation}
for $r = 20$, measured in m. 

In the following, we study different mechanisms of the model in more depth.

\subsubsection{Combustion function}

In the ADR-model~\eqref{eq:finalModel}, the combustion function $\Psi$ was still general and not defined in detail. 
For studying the influence of the combustion function, we do not consider for now any advection influence and set $\mathbf{v}=v=0$. 

\begin{figure}
    \centering
    \includegraphics[width = 7cm]{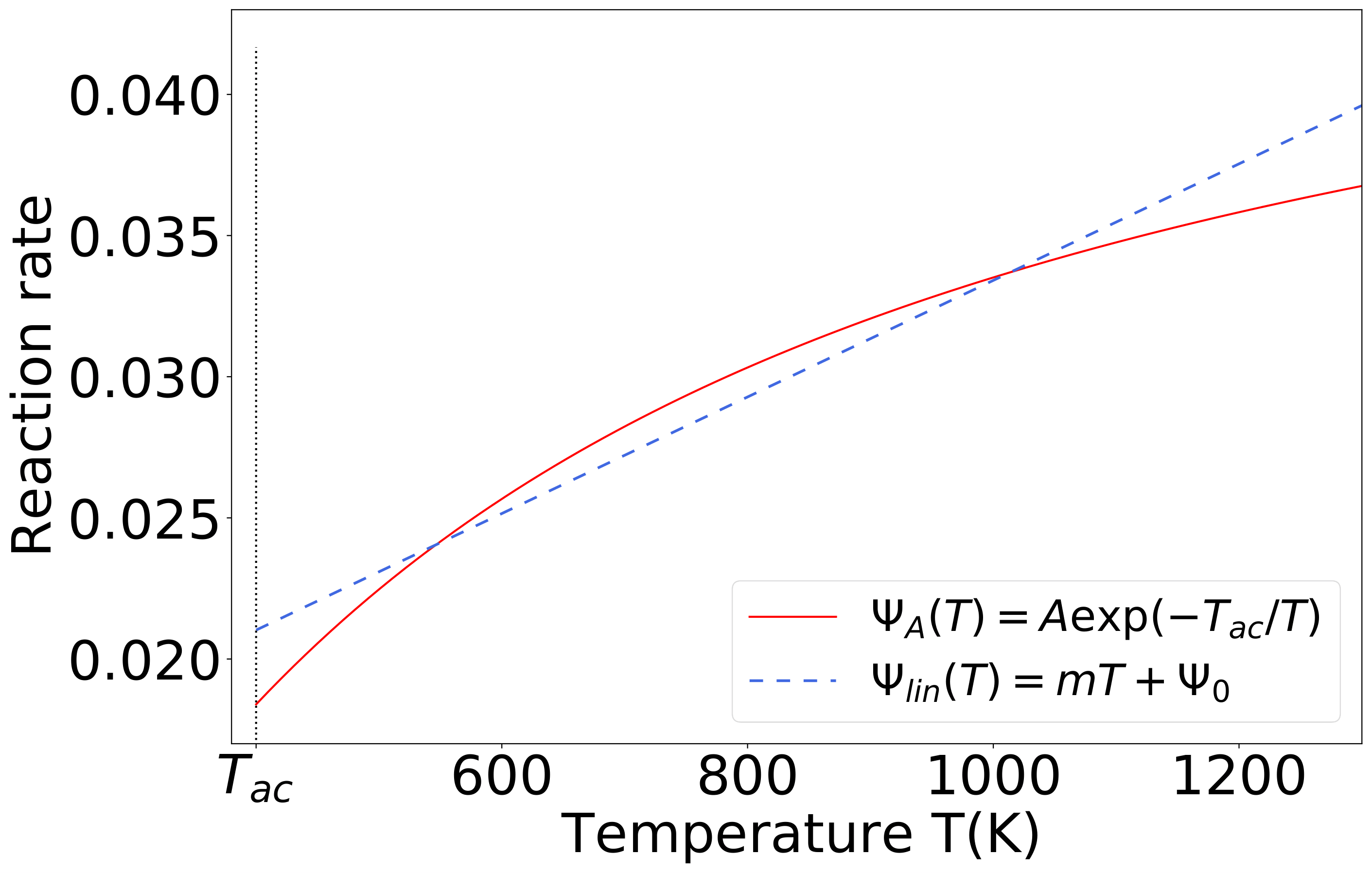}
    \caption{ 
    Reaction rates governed by the Arrhenius law $\Psi_\mathrm{A}(T)$ and its linear approximation $\Psi_\mathrm{lin}(T)$ in the relevant temperature domain.}
    \label{fig:solArrhenius1}
\end{figure}

Based on chemical reactions, the combustion process is traditionally modelled with the Arrhenius law, see \cite{asensio_wildland_2002}, giving 
\begin{equation}
    \Psi_A (T) = s(T) A \exp \left (-\frac{T_\mathrm{ac}}{T} \right),
    \label{firstPsi}
\end{equation}
with the activation function 
\begin{equation}
    s(T) = \left\{\begin{array}{ll} 0,  & T<T_\mathrm{pc}\\
         1, & T \geq T_\mathrm{pc}. \end{array}\right.
\end{equation}
that describes that the combustion starts above a threshold $T_\mathrm{pc}$.
Fig.~\ref{fig:solArrhenius1} shows the reaction rate $\Psi_A $ in the relevant temperature domain. 

As the Arrhenius law is formulated for molecular reactions and models the reaction process on our scale with an alleged precision, we propose a linear approximation instead. 

Through a least square fitting in the relevant temperature domain $T\in [T_\mathrm{ac}, T_\mathrm{max}] $ with $T_\mathrm{max}=1200\,$K, we find a linearized combustion function 
\begin{equation}
    \Psi_\mathrm{lin}(T) = s(T)(m T + \Psi_0 ), 
        \label{eq:linArrhenius}
\end{equation}
with $m = 2.0646 \times 10^{-5}\, \mathrm{K}^{-1}$ and $ \Psi_0 = 0.0128\,$K being determined by a least squares approximation.

\begin{figure}[hbt]
    \centering
    \includegraphics[width = 7cm]{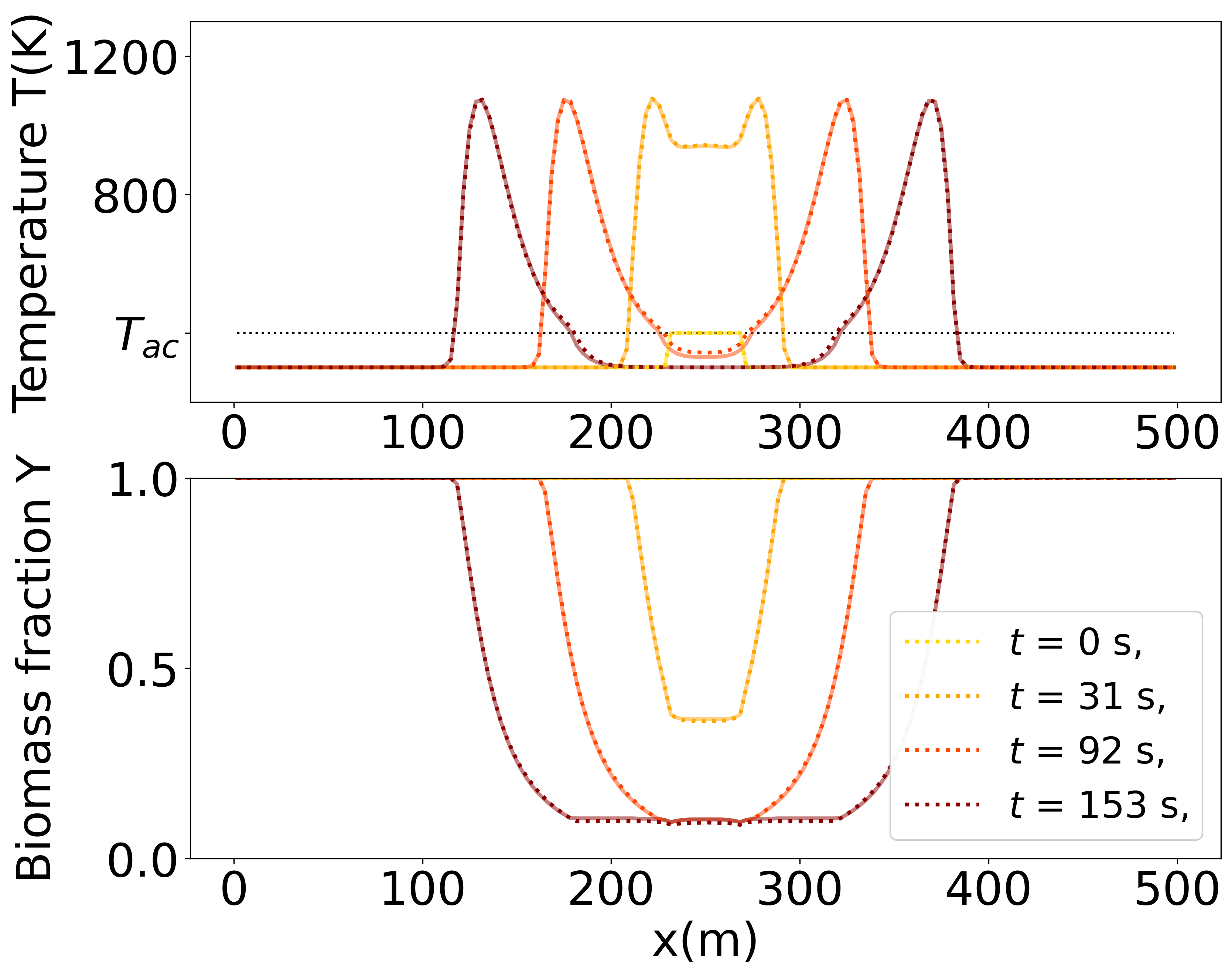}  
    \caption{ 
   Numerical solution of the ADR-model \eqref{eq:finalModel} with combustion functions following the Arrhenius law $\Psi_\mathrm{A}(T)$ \eqref{firstPsi} (dotted line) and its linear approximation $\Psi_\mathrm{lin}(T)$ \eqref{eq:linArrhenius} (solid line). 
      The simulation results overlap almost everywhere identically.}
    \label{fig:solArrhenius}
\end{figure}

Fig.~\ref{fig:solArrhenius} shows the solutions for both reaction functions. 
In general, the temperature forms travelling wave fronts, modeling the fire spread. The temperature decreases due to convection after the combustion process eliminates most of the biomass. 
The spread of the travelling wave speed is caused by the interplay of diffusion and reaction. 

The comparison of the travelling wave solutions for the two reaction rates in Fig.~\ref{fig:solArrhenius} shows only slight differences in temperature below the activation temperature $T_\mathrm{ac}$. Besides, the speed of the travelling waves is almost identical, and so is the shape of the travelling wave front. 

Consequently, the linear reaction rate gives a good approximation for the Arrhenius reaction rate. 
The advantage of Eq.~\eqref{eq:linArrhenius} compared to \eqref{firstPsi} consists in the easier mathematical terms, so that the local reaction system resembles a Lotka-Volterra system. Furthermore, Eq.~\eqref{eq:linArrhenius} can be regarded as a representative of the class of possible combustion functions under uncertain knowledge.

\subsubsection{Advection velocity} 
In the previous study, we set the advection velocity to $v = 0\,$ to gain a better understanding of the impact of different combustion functions. Given the crucial role of wind in fire propagation, this study explores the impact of varying advection velocities on the model's dynamics. 
We assume a linear connection $v= \beta w $ between the advection velocity $v$ and the wind velocity $w$.

\begin{figure}
\centering     
\subfigure[]{\label{fig:winda}\includegraphics[width=92mm]{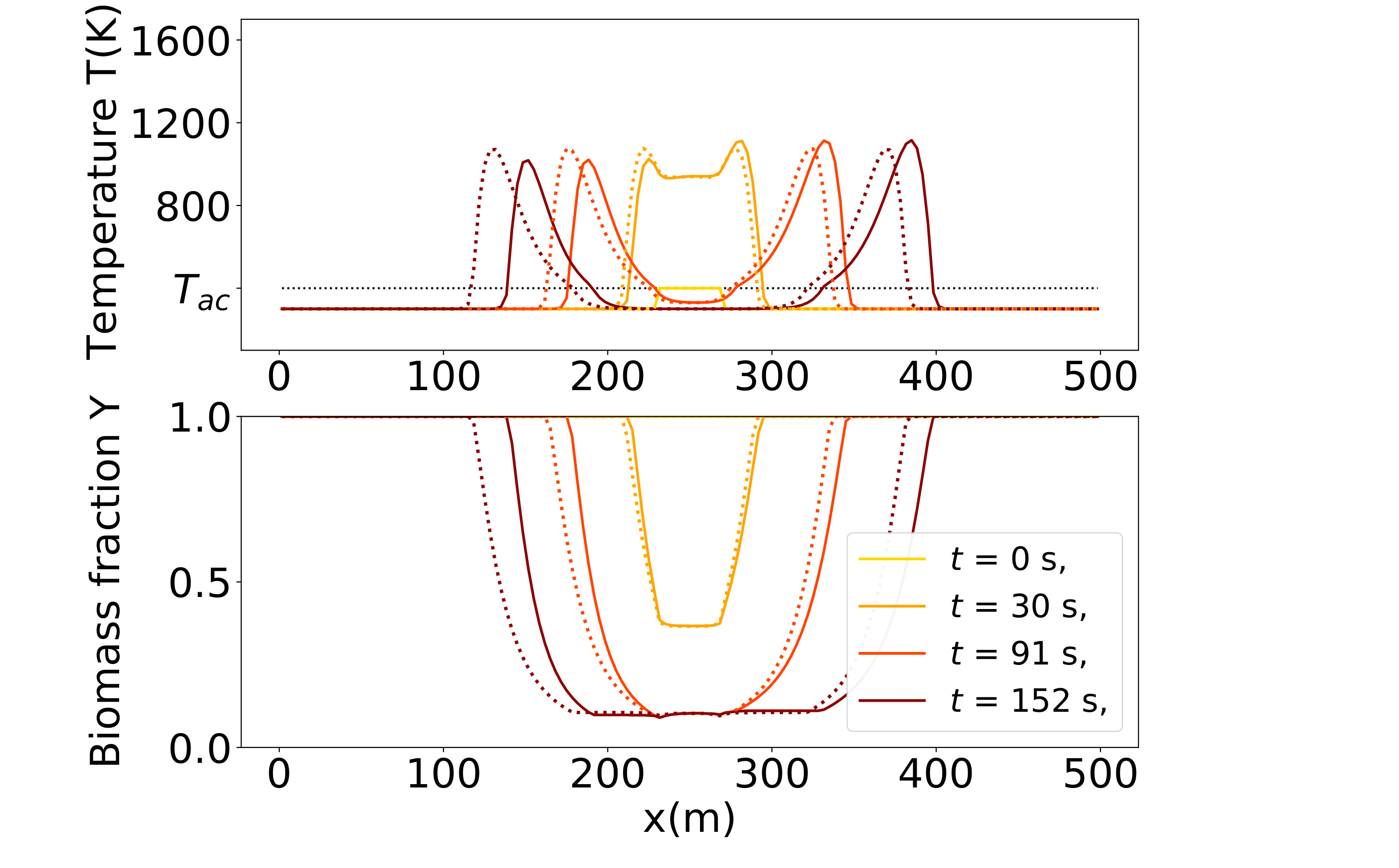}
}
\subfigure[]{\label{fig:windb}\includegraphics[width=92mm]{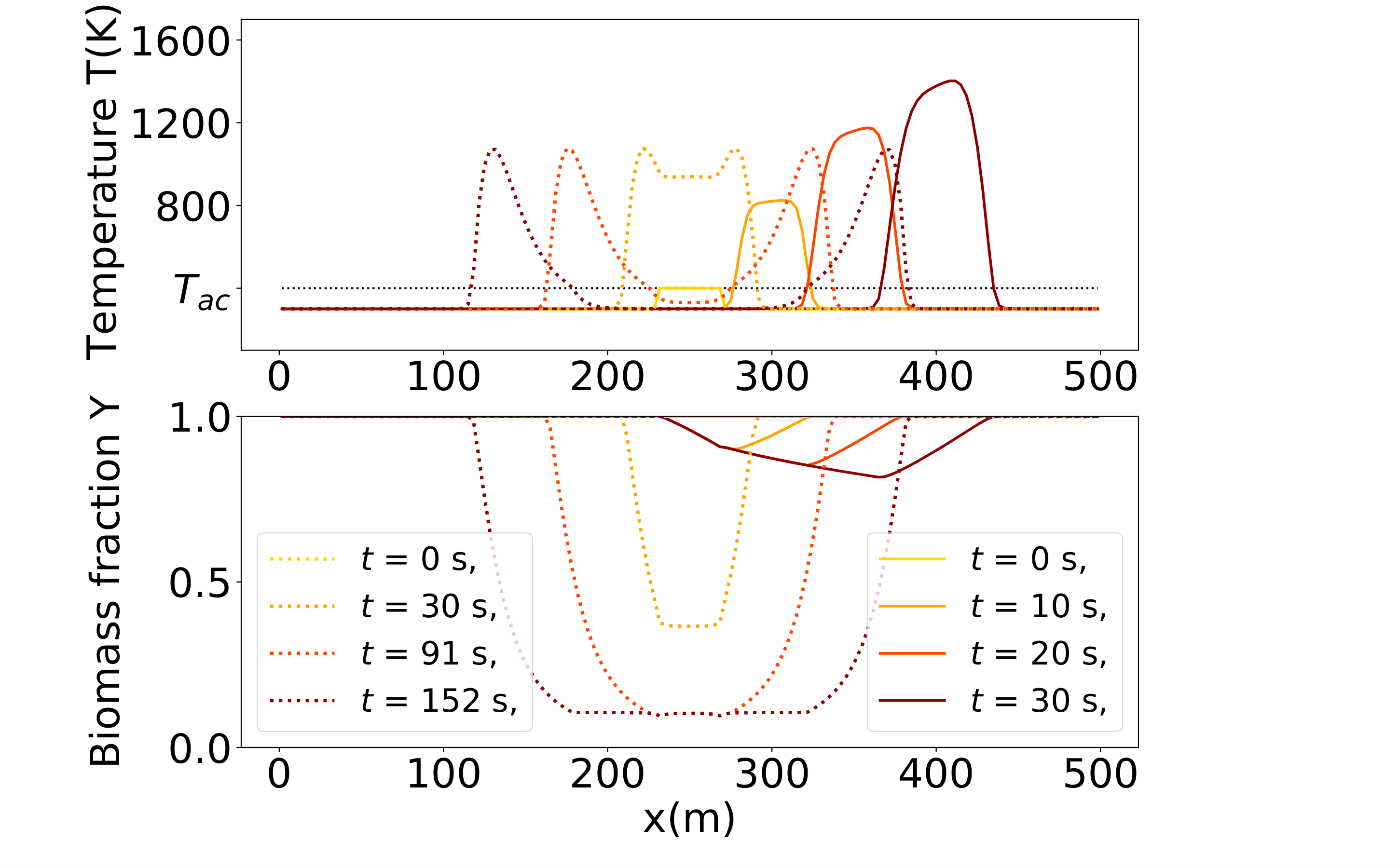}}
\caption{ Numerical solutions of the ADR - model \eqref{eq:finalModel} 
without wind (dotted line) and with wind in $x$-direction (solid line) with advection velocities $v = 0.1$\,m/s (upper) and $v~=~10.0$\,m/s (lower).}
\label{fig:wind}
\end{figure}

In Fig. \ref{fig:wind} we present two scenarios: one with a very low velocity $v = 0.01\,$m/s (Fig. \ref{fig:winda}) and another with a more realistic velocity $v = 10.0\,$m/s (Fig. \ref{fig:windb}), extracted from \cite{fire_velocity}.  In both cases, there is  noticeable transport of temperature in the wind direction, with the transport speed increasing as the advection velocity rises. From Fig. \ref{fig:wind} (lower) we observe, that the heat transport becomes a dominant process at higher velocities, leading to strongly increased maximum temperatures in the wind direction.
Additionally, we see that the biomass consumption decreases for high wind and therefore high advection velocities. 

This study demonstrates the substantial influence of wind on the system's dynamics. 
Additionally, the advection speed is modified by uneven topography, which in turn can affect the direction of fire propagation due to variations in slope. Therefore, it is instructive to investigate the impact of topography by incorporating it into the advection model.

\section{Topography extension of the ADR - model}
In this section, we introduce a model extension to the ADR-model incorporating topography as an additional advective term. We examine the impact of topography on the dynamics using basic sample topographies in one and two dimensions. Afterward, we discuss the limitations of the models and propose potential corrections. 

\subsection{Topography-expansion}\label{sec:topography}

 Since this study aims to examine the broader effects of topography on fire front propagation rather than the fluid dynamics of varying topographies, we neglect complex fluxes in our approach. 
Experiments show that the propagation velocity increases with steeper terrain inclines, see \cite{sanchez-monroy_fire_2019, abouali_analysis_2021}.
 This direction-dependent acceleration can be described by a virtual wind created by the topography. Therefore, we assume that the virtual wind is proportional to the topography gradient, as previously examined by \cite{burger_exploring_2020}. We redefine the advection velocity $\mathbf{v}$ as 
\begin{equation}
    \mathbf{v} = \beta \mathbf{w} + \gamma \nabla Z(\mathbf{x}),
\label{eq:virtualWind}
\end{equation}
where $\mathbf{w}$ is the natural wind in the area, $Z(\mathbf{x})$ is the topography, and $\beta$ and $\gamma$ are calibration factors.
 We incorporate this new advection velocity into the ADR-model \eqref{eq:finalModel}, which leads to the modified differential equations
\begin{align}  
\begin{aligned}
    \hspace{.2cm} \rho_0 c & \left(\frac{\partial T}{\partial t} + \nabla \cdot (T (\beta \mathbf{w}+ \gamma \nabla Z))  \right) \\ &=  k \Delta T - \alpha (T-T_\infty) + \rho_0 \mathcal{H} \Psi(T) Y  , \\ 
         \hspace{.2cm}  \frac{\partial Y}{\partial t} & =  - \Psi(T) Y.
\end{aligned}
    \label{eq:TopoModel}
\end{align}
Given the significant impact of the natural wind on the direction of the fire front propagation, as discussed previously, we set $\mathbf{w} = \mathbf{0}$ for further analysis to isolate and study the effect of the topography. Therefore, we also set both weighting factors to $\beta = \gamma = 1$.

\begin{figure}
    \centering
    \includegraphics[width = 8cm]{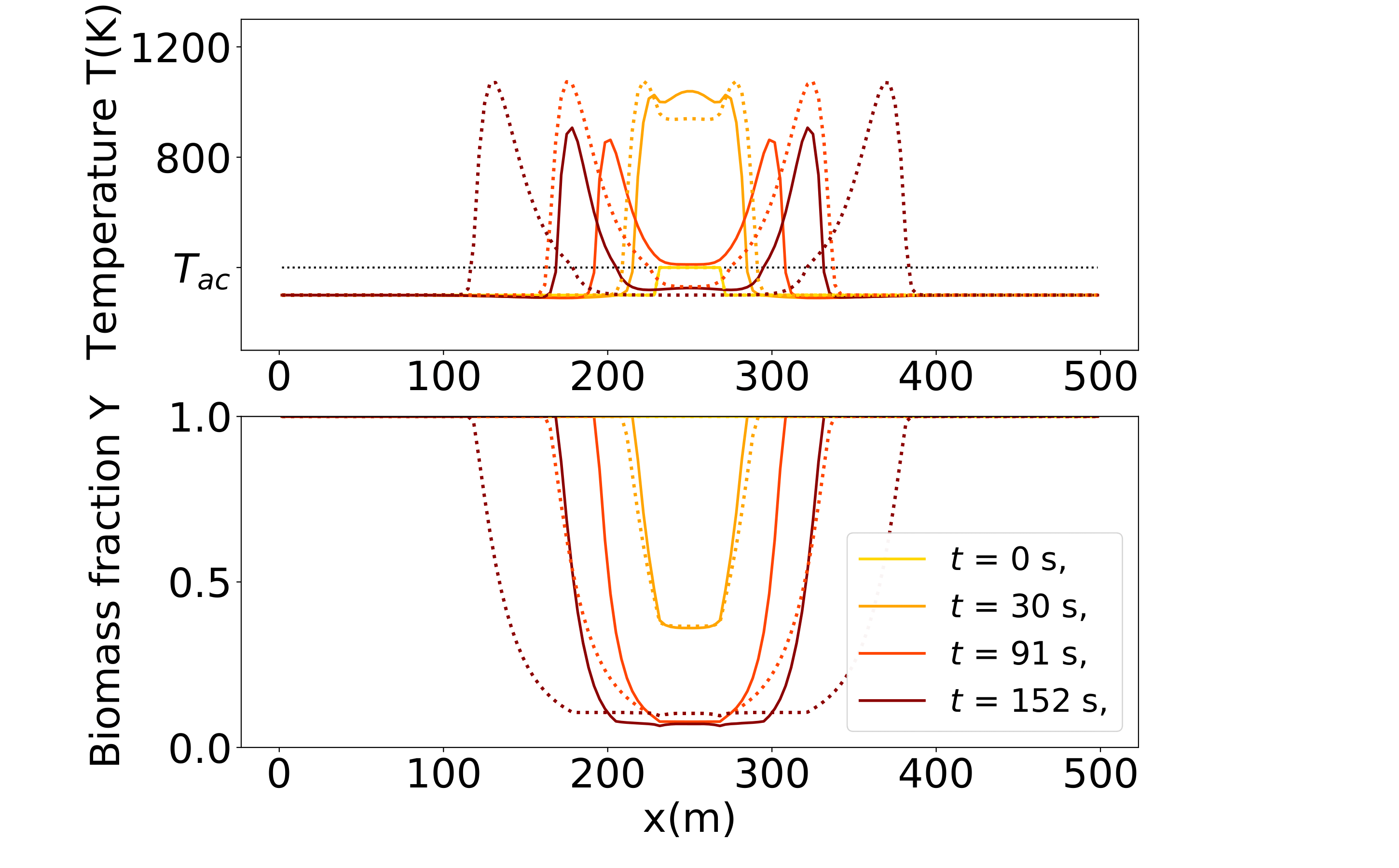}
    \caption{Numerical solutions of the ADR - model \eqref{eq:finalModel} (dotted line) and the topography model \eqref{eq:TopoModel} with topo\-graphy \eqref{eq:testTopography} resembling a hill centered at $x_0~=~250\,$m (solid line). 
}
    \label{fig:comparisonADR-Topography}
\end{figure}

\subsection{Dynamics of the ADR-model with topography}
 To understand the influence of topography on the overall behavior of the system, it is beneficial to study the dynamics with straightforward topographies. 
 \subsubsection{Dynamics with sample topographies}
 We choose a simple one-dimensional topography
\begin{equation}
    Z(x) = Z_0 \exp\left(-\frac{(x-x_0)^2}{4000}\right),
    \label{eq:testTopography}
\end{equation}
resembling a hill centered at $x_0 = 250\,$m in the area with a height $Z_0 = 20\,$m. 
In Fig.~\ref{fig:comparisonADR-Topography} the comparison of the modified model \eqref{eq:TopoModel} to the ADR-model \eqref{eq:finalModel} is shown. The model incorporating topography displays similar overall behavior but exhibits higher temperatures around the center $x_0 = 250\,$m of the area, i.e. on the top of the hill, while the travelling wave speed is reduced. The virtual wind, directed to the center, retains heat longer in the vicinity of $x_0$.
\begin{figure}
    \includegraphics[width=.24\textwidth]{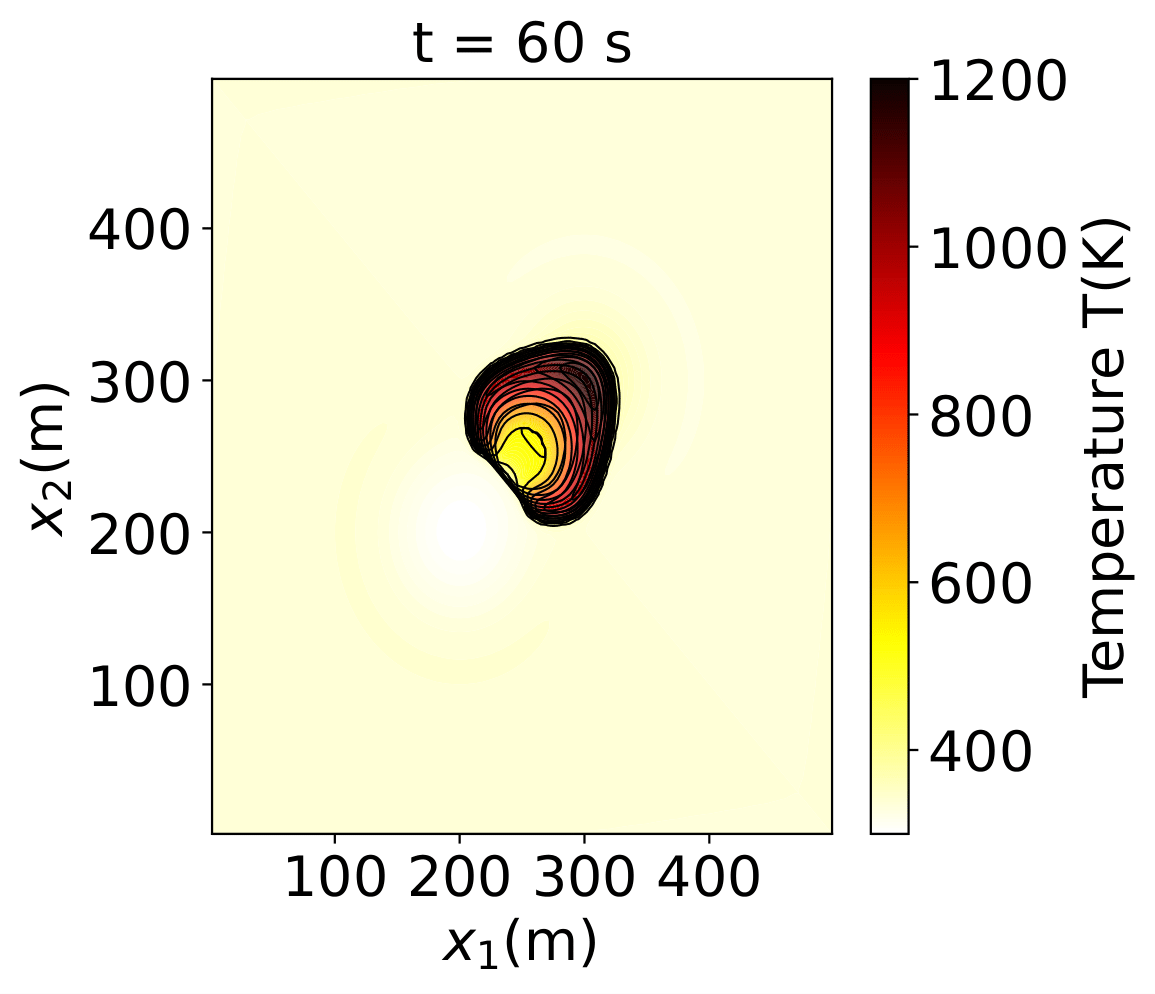}\hfill
    \includegraphics[width=.24\textwidth]{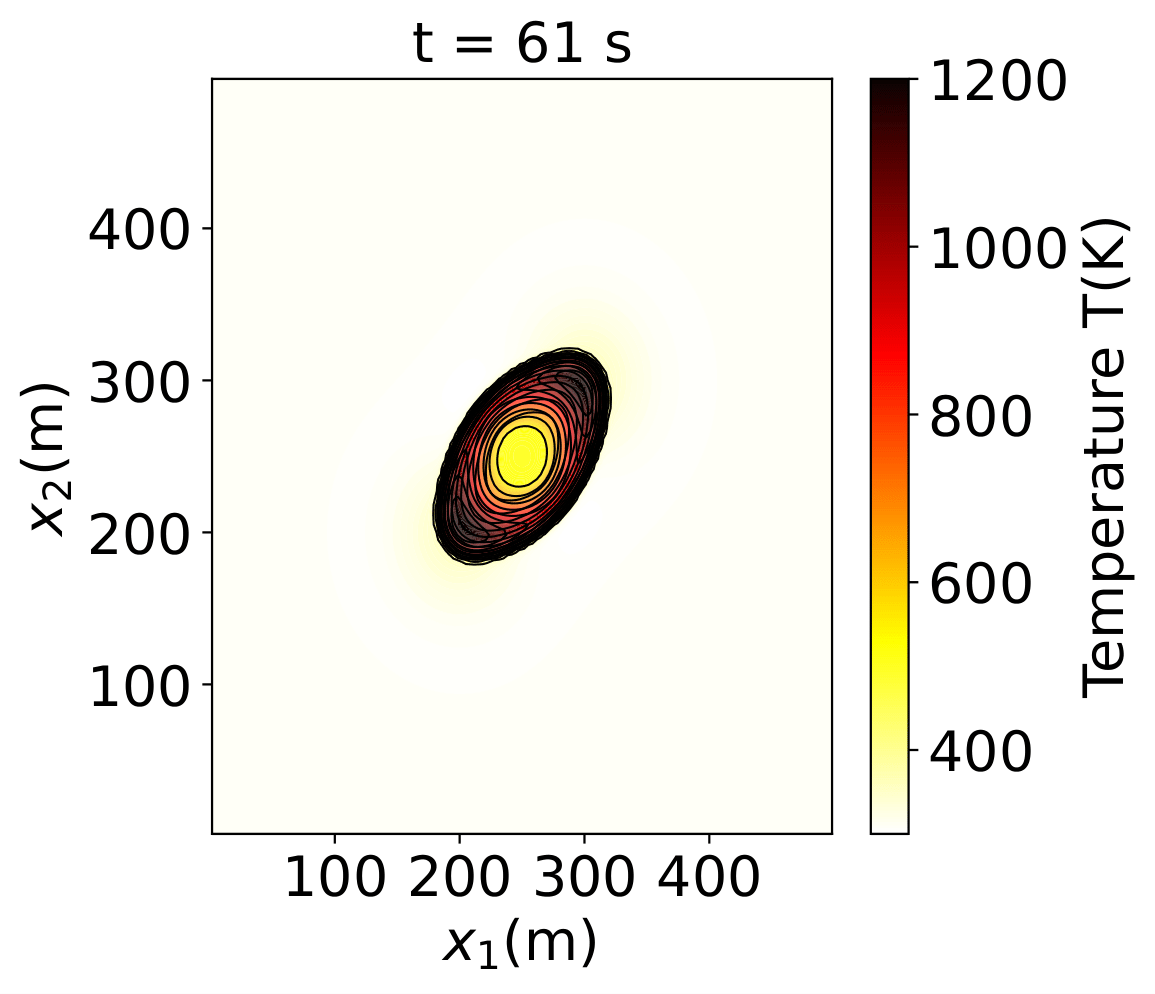}
    \\[\smallskipamount]
    \includegraphics[width=.24\textwidth]{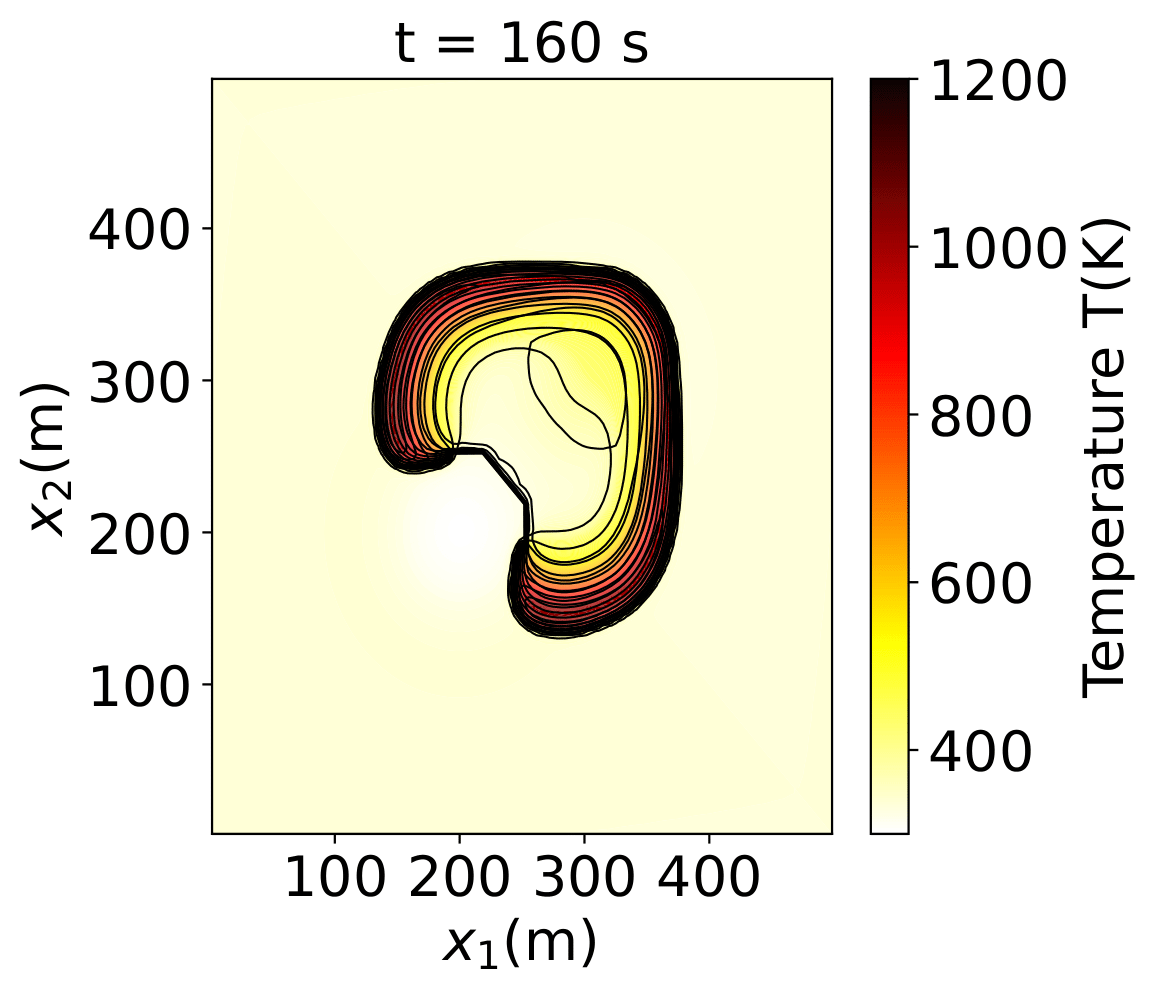}\hfill
    \includegraphics[width=.24\textwidth]{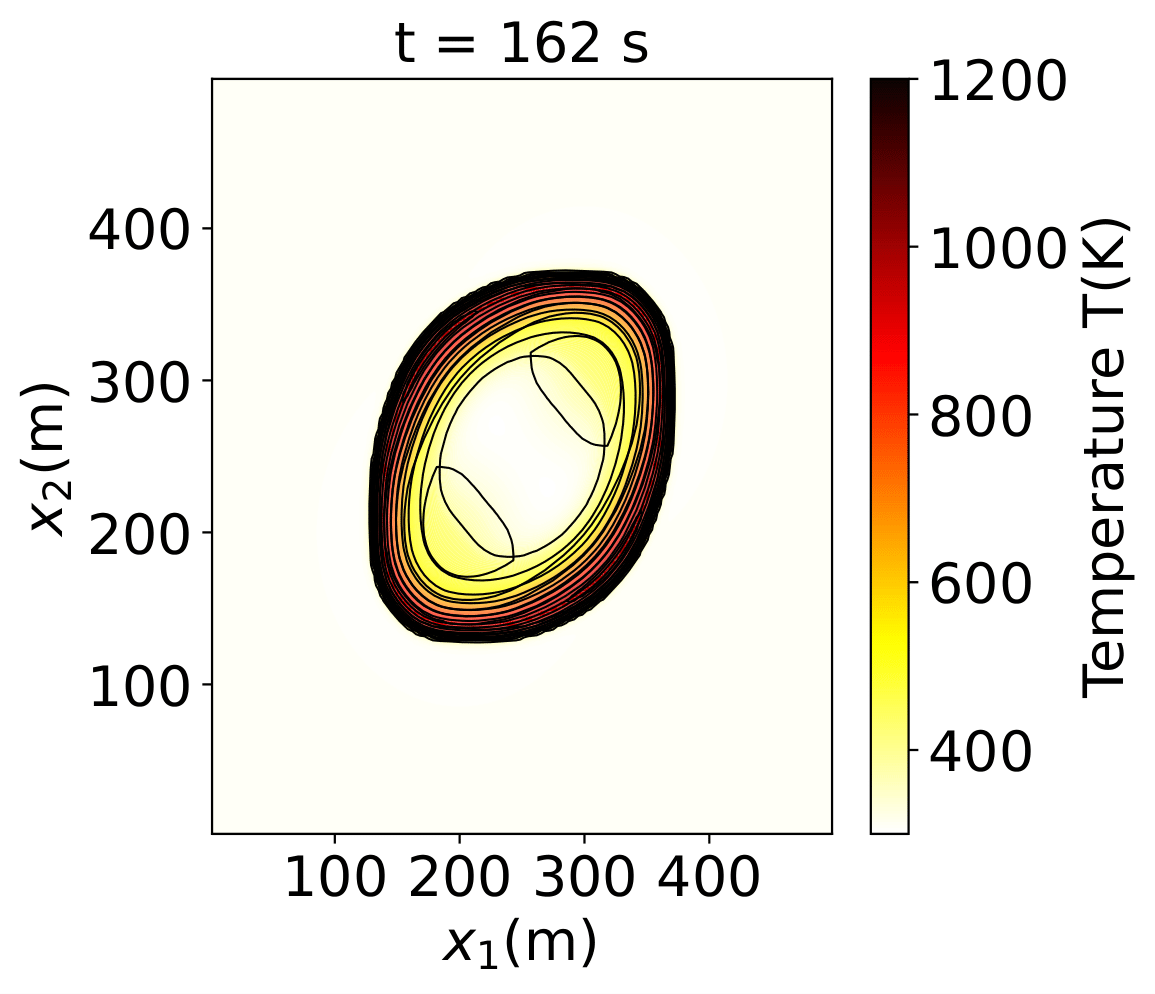}
    \caption{Fire front propagation in two dimensions at two time steps with the topography \eqref{eq:2DTopography} with different heights. (Left) Solution for a valley with height $Z_1 = -20\,$m at $\mathbf{u}_1 = (200,200)^\mathrm{T}$ and a hill with height $Z_2 = 20\,$m at $\mathbf{u}_2 = (300,300)^\mathrm{T}$. (Right) Solution for two hills with heights $Z_1 = Z_2 = 20\,$m each centered at $\mathbf{u}_1$ and $\mathbf{u}_2$.}
    \label{fig:2Dtopography}
\end{figure}

Given that wildfires spread in two dimensions, it is instructive to evaluate the impact of topography in a two-dimensional context. Extending the previous topography, we model the simple cases of a hill next to a valley, and two hills by 
\begin{equation}
\begin{split}
    Z(\mathbf{x}) &= Z_1 \exp(-\lVert \mathbf{x}-\mathbf{u}_1\rVert^2/4000)\\
    &+ Z_2 \exp(-\lVert\mathbf{x}-\mathbf{u}_2\rVert^2/4000),    
\end{split}
    \label{eq:2DTopography}
\end{equation}
shown in Fig. \ref{fig:2Dtopography}, where $\mathbf{u}_1 = (200,200)^\mathrm{T}$, $\mathbf{u}_2 = (300,300)^\mathrm{T}$, $-Z_1 = Z_2 = 20\,$m (left) and $Z_1 = Z_2 = 20\,$m (right).  
From the figures, we observe a strong dependence of the fire front shapes on the topography. In the scenario of a valley and a hill (Fig. \ref{fig:2Dtopography}, left), the fire advances uphill, while the temperature in the valley remains nearly constant at the initial temperature of $T_0 = 300\,$K. In the process, the fire spreads along the edges of the valley. Conversely, in the scenario with two hills, the fire spreads towards both hills, creating an elliptical fire front shape.

\subsubsection{Heat accumulation and corrections}
\begin{figure}[t]
\centering
    \includegraphics[width = 4.2cm]{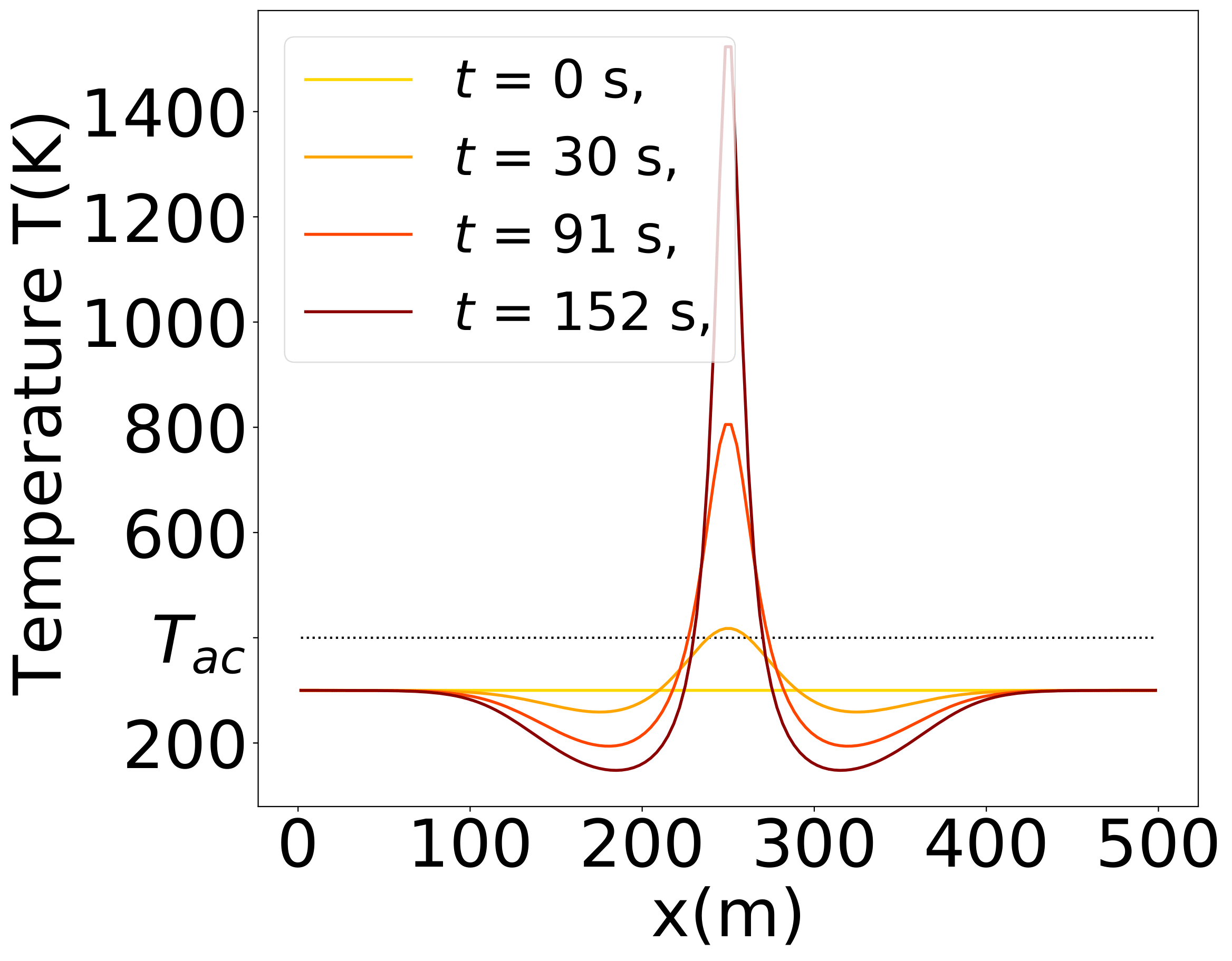}
    \includegraphics[width = 4.2cm]{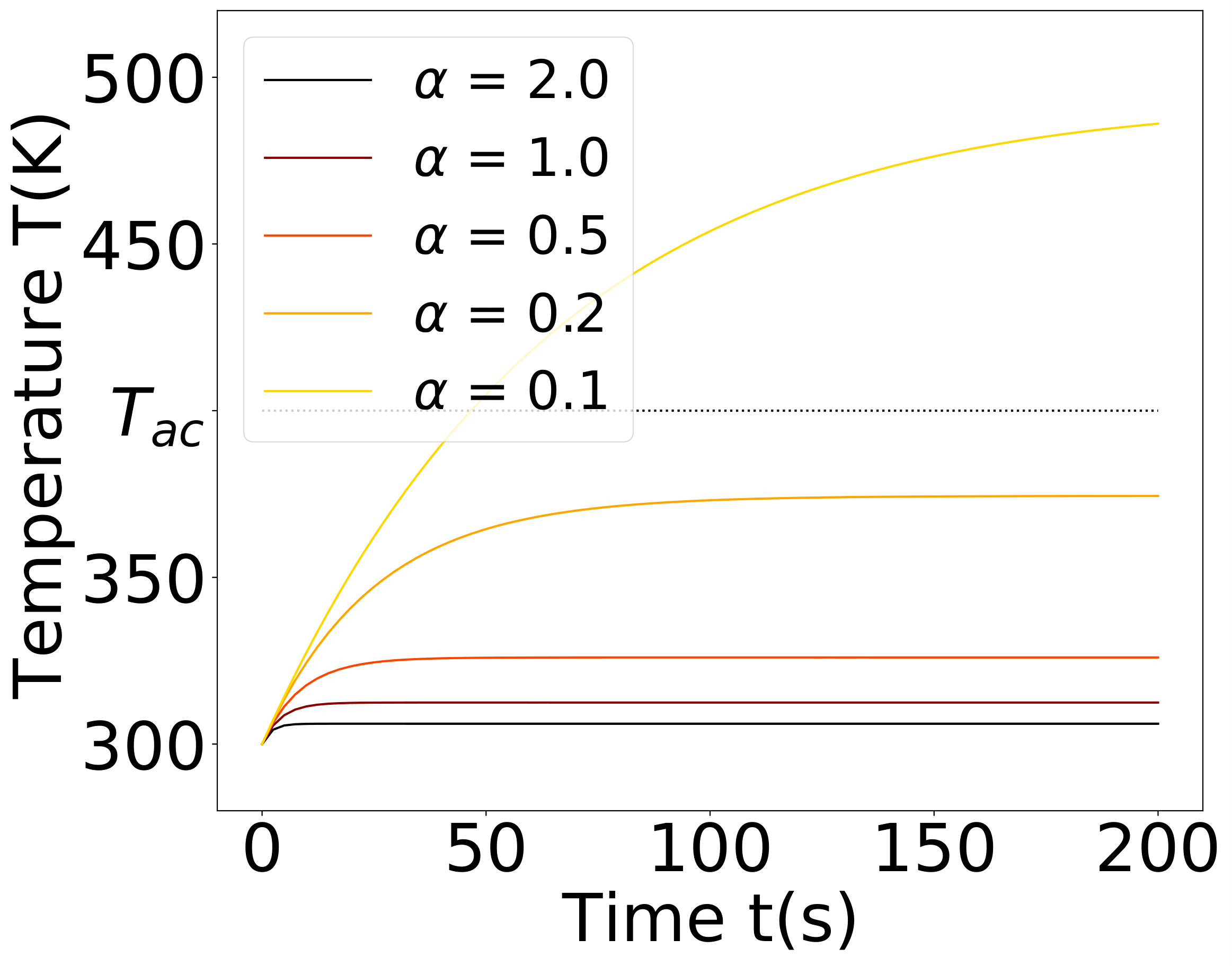}%\\
    \caption{(Left) Numerical solution of the ADR-model with constant initial temperature ($T_0(\mathbf{x})=300\,$K) with only the advective term using topography \eqref{eq:testTopography}. (Right) Time evolution of the temperature~$T(x_0,t)$ in the middle of the area with only the convective and advective topography term for different coefficients of heat transfer $\alpha$, which governs the impact of cooling due to convection.}
    \label{fig:heatAccumulation_top}
\end{figure}

Although the model demonstrates the expected dynamics for added topographies, it faces an issue with energy conservation. Due to the two-dimensionality of the model, the virtual wind exhibits a nonzero divergence, leading to a heat accumulation. Fig.~\ref{fig:heatAccumulation_top}~(left), shows the temperature evolution for the simple topography \eqref{eq:testTopography} considering only the advective term in the model \eqref{eq:TopoModel} ($\alpha = k = \mathcal{H} = 0$) with a constant initial temperature $   T_0(\mathbf{x}) = 300\,\mathrm{K}, \, \, \mathbf{x} \in \Omega.$
Despite this constant initial temperature, the temperature in the center of the area rises rapidly, surpassing the activation temperature $T_\mathrm{ac}$ purely due to advection, independent of reaction. This shows that a possible ignition of fire can occur due to advection, which is unrealistic.

We propose two approaches to correct this heat accumulation.
 First, it is instructive to examine the interaction of the other mechanisms already included in the model~\eqref{eq:TopoModel} with this heat accumulation. Newton's cooling law \eqref{eq:KonvQ} works as a counteracting term to the temperature increase, as displayed in Fig.~\ref{fig:heatAccumulation_top}~(right). The temperature time evolution shows convergent behavior with the maximum value rising as $\alpha$ increases. Thus, by appropriately adjusting $\alpha$ the heat accumulation can be mitigated.

 The second approach cancels the nonzero divergence of the virtual wind by introducing a vertical flow $\Tilde{w}$, called topflow, to the model. This topflow aims to correct the non-compressibility of the advection $\mathbf{v}$ \eqref{eq:virtualWind} by adding a wind in the vertical direction, allowing the accumulated heat to escape. We define this wind by the property of a total non-zero divergence 
\begin{equation} 
     \nabla \cdot \begin{pmatrix} \mathbf{v} \\ \Tilde{w}\end{pmatrix} = 0,  
\end{equation}
for both, $\Tilde{w} \in \mathbb{R}^1$ and $\mathbf{v}\in \mathbb{R}^2$.
From this condition, we derive for the topflow
\begin{equation}
    \Tilde{w} = - \frac{1}{S} \int_{\partial V} \mathbf{v} \cdot \mathbf{n} \, \mathrm{d}\Gamma.
\end{equation}
Accordingly, we add the additional convective term  
\begin{equation}
    \dot{Q}'_\mathrm{conv} = \frac{\rho_0 c}{\ell} \int_S T(\mathbf{x},t) \Tilde{w}\, \mathrm{d}\Gamma,
    \label{eq:differentialQconv'}
\end{equation}
to the system, resulting in 
\begin{align}
\begin{aligned}
     \rho_0 c & \left(\frac{\partial T}{\partial t} + \nabla \cdot (T (\beta \mathbf{w}+ \gamma \nabla Z)) + T \Tilde{w}  \right)  \\ &=  k \Delta T - \alpha (T-T_\infty) + \rho_0 \mathcal{H} \Psi(T) Y ,  \\ 
         \hspace{.2cm}  \frac{\partial Y}{\partial t} & =  - \Psi(T) Y.
\end{aligned}
    \label{eq:TopflowModel}
\end{align}

\begin{figure}[b]
    \centering
    \includegraphics[width = 8cm]{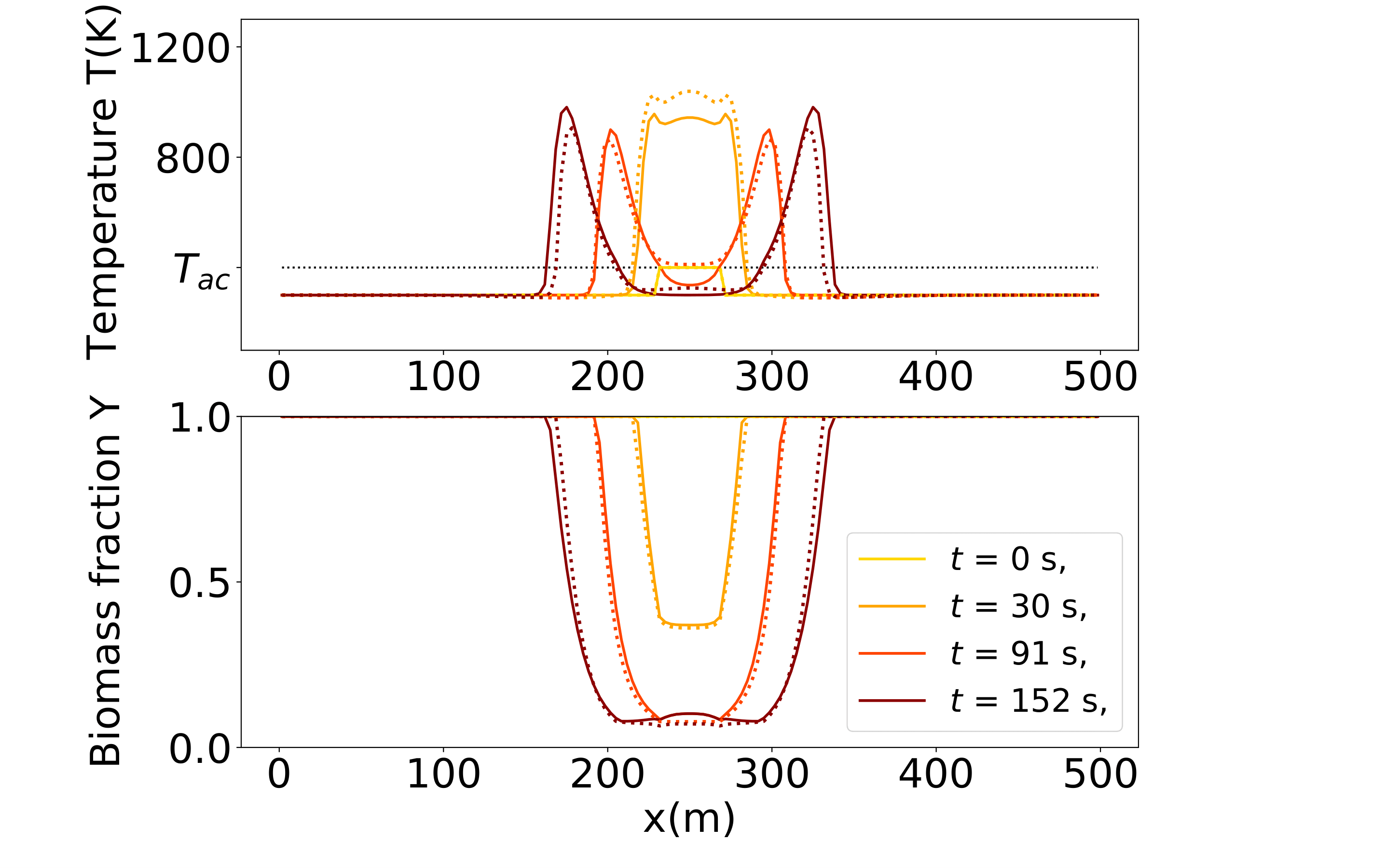}
    \caption{Solution of the ADR-model with topography \eqref{eq:TopoModel} (dotted line) compared to the model including an additional vertical flow  \eqref{eq:TopflowModel} (solid line).}
    \label{fig:heatAccumulation_bottom}
\end{figure}

Fig. \ref{fig:heatAccumulation_bottom} shows that the numerical simulation exhibits no heat accumulation due to topography while retaining the overall dynamics of the topography model \eqref{eq:TopoModel}.

Another possibility to circumvent the unnatural heat accumulation is to rewrite the advection term in non-conservative form, i.e. $\mathbf{v}\cdot \nabla T$ as done in \cite{burger_exploring_2020}. To derive this version of the model, incompressibility of the advection velocity field should be assumed. 

\section{Conclusion}

In a general modeling perspective, we constructed a model family where single mechanisms combine to different models. Every summand in Eq.~\eqref{eq:finalModel} is related to a mechanism like advection, diffusion or reaction, and the selection of the mechanisms is a first modeling technique used here. \\
The second modeling technique is the choice of simplest possible representatives, e.\,g.\ for the combustion function $\Psi$ in Eq.~\eqref{eq:linArrhenius}, which do not affect the qualitative solution behavior. It might seem counterintuitive to reject the accepted theoretical knowledge of the Arrhenius law in Eq.~\eqref{firstPsi}, but the acceptance belongs to chemistry and cannot be directly transferred to wildfires.\\
As a third modeling technique, the expansion of an existing model is presented, like the topography extension of the ADR-model in Sec.~\ref{sec:topography} and its necessary phenomenological correction by the hierarchically subsequent topflow summand in Eq.~\eqref{eq:TopflowModel}.\\
Besides the practical use in modeling wildfires, this model family is a nearly ideal example for a developing modeling of a life-science application with some uncertain knowledge. Since the models are still rather small, e.\,g.\ compared to a full CFD-simulation, some learning process of the parameters on the base of measurements adapts the model behavior closer to realistic observations.

\begin{ack}
The work was partially funded by the Ministerio de Ciencia e Innovación (Agencia Estatal de Investigación) under project-nr. PID2022-141051NA-I00. 
\end{ack}

\bibliography{ifacconf}             % bib file to produce the bibliography
                                                     % with bibtex (preferred)
                                                   
\end{document}